\def\erg{{\rm\thinspace erg}}

\def\MeV{{\rm\thinspace MeV}}

\def\K{{\rm\thinspace K}}

\def\Msun{\hbox{$\rm\thinspace M_{\odot}$}}

\def\s{{\rm\thinspace s}}

\def\ergps{\hbox{$\erg\s^{-1}\,$}}

\def\spose#1{\hbox to 0pt{#1\hss}}
\def\approxlt{\mathrel{\spose{\lower 3pt\hbox{$\sim$}}
        \raise 2.0pt\hbox{$<$}}}
\def\approxgt{\mathrel{\spose{\lower 3pt\hbox{$\sim$}}
        \raise 2.0pt\hbox{$>$}}}

\documentstyle[psfig]{mn}

\title
[Two-temperature coronae in active galactic nuclei]
{Two-temperature coronae in active galactic nuclei}

\author[]
{T.~Di Matteo, E.G.~Blackman and A.C.~Fabian\\ 
{Institute of Astronomy, Madingley Road, Cambridge, CB3 OHA}\\}

\begin{document}

\maketitle

\begin{abstract} 
We show that coronal magnetic dissipation in thin active 
sheets that sandwich standard thin accretion disks 
in active galactic nuclei may account for 
canonical electron temperatures of a few $\times 10^9$K
if protons acquire most of the dissipated energy. 
Coulomb collisions transfer energy from the 
ions to the electrons, which subsequently cool rapidly by 
inverse-Compton scattering.  In equilibrium, the proton energy density
likely exceeds that of the magnetic field and both well
exceed the electron and photon energy densities. 
The Coulomb energy transfer from protons to electrons 
is slow enough to maintain a high proton temperature, but
fast enough to explain observed rapid X-ray variabilities in Seyferts. 
The $\sim 10^9$K electron temperature is insensitive 
to the proton temperature when the latter is $\ge 10^{12}$K.

\end{abstract}


\section{Introduction}

UV and X-ray observations of active galactic nuclei (AGN),
particularly in Seyfert 1 galaxies, indicate that the gravitational
binding energy of a massive black-hole is dissipated partly in a cold
accretion disk and partly in a hot corona above it. Thermal
Comptonization of soft UV-radiation in the corona leads to the
production of a hard X-ray continuum, some of which is reprocessed by
the cold disk (Haardt \& Maraschi 1991, 1993). In order to explain the
different ratios of X-ray and UV luminosities in different objects, it
has been suggested that the corona consists of localized active
regions. It is likely that these are produced by magnetic fields in
the disk amplified through differential rotation.  When the disk
magnetic field builds up significantly, buoyancy forces the field out
and above the disk, giving rise to active regions of high magnetic
field.  Magnetic field buoyancy can maintain both magnetically and
thermally dominated disks (Galeev, Rosner \& Vaiana 1979, Shibata et
al. 1990; Field \& Rogers 1993).

Magnetic dissipation (for example in a reconnection site) may more effectively
energize protons  than the electron. Since it is the
electrons which radiate, it is therefore plausible that the corona has
two kinetic temperatures, that of the protons, $T_{\rm p}$ and that of the
electrons, $T_{\rm e}$. These temperatures depend crucially on the
energy exchange rate from protons to electrons.


In this Letter we show that an optically thin 
corona above a thin disk can generate the appropriate 
electron temperature to explain the
observations of the hard continuum of AGN if electrons and protons are
coupled via two-body Coulomb collisions. 
Two-temperature models are usually considered in the context
of low density ion supported tori and advection
dominated thick disks (e.g. Shapiro, Lightman and Eardley 1976;  
Rees et al., 1982; Narayan and Yi 1995).  
In our context however, although the thin magnetically active coronal sheets 
are much less dense than their underlying 
thin dense disk, they can still have a much higher density than 
standard advection dominated disks.  We show that
this enables enough energy transfer by Coulomb electron heating
to satisfy constraints of rapid X-ray variability observed in many Seyfert I
nuclei (e.g. Zdziarski et al 1994, 1995).  Also, the active region need 
not be infalling, but can in fact be wind or ejection 
dominated and still transfer the required energy to electrons. 

To obtain this solution,  we balance
the energy transfer rates from the magnetic field to 
the protons and from the electrons to radiation. 
Balancing  these rates does not presume that any of the
component energy densities are equal.  We find that
the proton and magnetic energy densities are both 
much larger than  equilibrium electron and photon energy densities.

\section{Magnetic Energy Dissipation}


Magnetic dissipation sites (e.g. reconnection and/or shocks)
in astrophysical plasmas always involve
large gradients in the magnetic field, and the presence of length
scales much smaller than the field gradients of the non-dissipating
regions.  This is because the induction equation governing the
magnetic field
\begin{equation}
\partial {\bf B}/\partial t=\nabla\times ({\bf v}\times{\bf
B})+\nu_m\nabla^2{\bf B},
\end{equation} where $\nu_m$ is the magnetic
diffusivity, implies flux freezing and no dissipation as long as the
last term is small.  The ratio of this last term to the second is the
magnetic Reynolds number $R_m= vL/\eta>>1$, where $L$ and $v$ are the
magnitude of velocity and gradient length scales.  The large sizes and
conductivities in AGN (and astrophysical systems generally) make
dissipation negligible except in regions of large field gradients,
ie. thin current sheets (e.g. Parker 1979).

The more pronounced the variability of the
AGN source, the more likely the fewer the number of coronal 
dissipation regions: 
if $N$ dissipation regions operate at a time, then a fluctuation
changing the number of dissipation regions, would change the luminosity
$\sim (N \pm n)/N$, where $n$ is the change in number of
regions.  If $N>>1$, $n\sim N^{1/2}$ so that the effective change in
luminosity would be small compared to when $N\sim 1$.

A natural site of coronal dissipation by reconnection
and its slow shocks, would be at the interface
between the newly rising field from the disk and any pre-existing
coronal field (Begelman 1990; Shibata et al. 1990).  Disspation
between pre-existing loops has also been considered (Field \& Rogers
1993).  In general, 
the dissipation
layer could be expected to cover a sizable cross sectional fraction of
the underlying disk.

The development of small turbulent length scales
in the magnetically active regions enables the transfer of
field energy into particles (e.g. Larosa et al. 1996). The turbulence
generates a cascade of magnetic scales and the true dissipation scale is
that below which the energy drains into particles rather than
into smaller magnetic structures.  The transfer of magnetic to
particle energy likely occurs from
MHD wave-particle interactions, for example by stochastic or shock Fermi
energization (e.g. Achterberg 1987; Eilek \& Hughes 1991; Larosa et
al., 1996).  Such energization
requires two types of particle scattering, both of which likely
favor protons for AGN coronae as we now discuss.

Particle scattering by MHD waves is the first required 
mechanism. The MHD waves are essentially magnetic
compressions on scales much larger than the particle gyro-radii
which can ``mirror'' particles as they approach. 
For stochastic Fermi energization, the scatterers move
in random directions at $\sim v_A$, the Alfv\'en speed.
Only the ``head-on'' collisions  between particles and scatterers 
transfer energy to the particles.
For particles moving faster than or of order $v_A$, 
both head-on and trailing scatterings occur.  
Even though much slower particles do see only head-on
collisions, there is a critical velocity perpendicular to the magnetic
field required by particle in order for them to be scattered (Larosa 
et al. 1996).
This latter condition dominates, and more
energy is gained for faster particles.  If the protons have
 higher velocities to begin with, then proton rather
than electron scattering would be favored.

The above scattering can only enhance a particle's momentum component
along the field.  In order to be repeatedly scattered and gain
significant energy, the particles must be pitch angle isotropized
with respect to the local magnetic field.  This pitch angle scattering
is accomplished by resonance with plasma Alfv\'en waves on the scale of
the particle gyro-radii (Achterberg 1987; Eilek \& Hughes 1991).  Such
waves only exist below a critical frequency and so the particles must
have significantly large gyro-radii before gyro-resonance can occur.
The required pitch angle scattering also favors the protons: in a
plasma for which the Alfv\'en speed is of order $c$, the electrons
need to be quite relativistic to resonate with Alfv\'en waves.  Such
favoring of protons has been called the ``electron injection''
problem (Eilek \& Hughes 1991).

Shock Fermi energization may also operate across slow shocks in the
reconnection region (Blackman \& Field 1994).  This
 produces only head- on collisions and is thus more
efficient than stochastic processes.  However, similar
arguments to those above would apply, and protons would be more easily 
energized.

The previous discussion suggests that 
when the protons are already at 
higher kinetic temperatures than the electrons,
magnetic dissipation can maintain such a state in equilibrium.  
Setting up the initial configuration is a separate problem
since the initial plasma might not necessarily have 
protons at a higher temperature than electrons.
However, even if the initial dissipation went into electrons, they would
cool very fast.  The proton energy density could slowly grow. 
After some time, the protons could acquire enough energy
to dominate the electrons, leading to the two-temperature equilibrium
we describe herein. We leave a more detailed investigation for future work.

Simulations (Priest \& Forbes, 1986)
as well as solar (Tsuneta 1994) and  the geomagnetic tail
observations  (Coroniti et al. 1994) 
indicate that magnetic reconnection occurs in the presence of slow
shocks.  This suggests that Petschek (Petschek 1962) type
reconnection can occur in nature, distinctly from Sweet-Parker (Sweet 1958;
Parker 1979) type reconnection.  The main difference between the two
for non(or mildly)-relativistic Alfv\'en speeds is that for 
the latter, the ratio of height
to width of the the overall dissipation region satisfies is $h/r \sim
v_{\rm dis}/V_A\sim 1/R_m^{1/2}$ while in the former the ratio is $h/r
\sim v_{\rm dis}/V_A \sim 1/log_{\rm e}[R_m]$ where the $R_m>>1 $ is measured
outside of the dissipation region.  The Sweet-Parker dissipation
region is generally thinner and magnetic energy conversion occurs much
slower.
Our calculations below, combined with
observational constraints, seem to suggest that $h/r\sim 1/10$ is
reasonable, possibly suggesting that the dissipation regions are
Petschek-like.

Finally, note that the processes discussed above
might generate non-thermal proton
distributions. However, the electron
energization times are also of order their Maxwellian relaxation times
(Spitzer 1956). Thus the electrons can be rapidly 
thermalized even if the protons are initially non-thermal.

\section{A two-temperature corona}
\subsection{Time scales}

We now motivate the model by exploring relevant time scales using
simple approximate formulae, reserving a fuller treatment for the next
Section.

Thermal Compton scattering gives the power-law shape to the observed hard
X-ray spectrum and exponential cut-off of Seyfert 1 nuclei, provided that
the plasma is optically thin to electron scattering (i.e. $\tau \approxlt
0.4$) and has an electron temperature $kT_{\rm e}\sim2-5\times 10^9\K$.
The Compton cooling time scale
\begin{equation}
t_{\rm cool}\approx \pi t_{\rm cross}/\ell
\end{equation}
where the compactness parameter $\ell={L/ r}({\sigma_{\rm
T}/{m_{\rm e} c^3}}).$ $L$ is the luminosity of the active region
which is of size $r$. For a source such as MCG--6-30-15 which has a
luminosity of $10^{43}\ergps$ and varies typically on a time scale of
$1000\s$ then $\ell\sim 10$, so $t_{\rm cool}\sim t_{\rm cross}/3,$ where
$t_{\rm cross}=r/c.$ The electrons cool on less than a light crossing
time.

The electron-proton coupling time, $t_{\rm ep}$ is defined from the
energy transfer equation (Spitzer 1956) 
\begin{eqnarray}
{{dT_{\rm
e}}\over{dt}}&=&{{(T_{\rm p}-T_{\rm e})}\over t_{\rm ep}}. \nonumber \\
t_{\rm
ep}&=&\sqrt{\pi\over2}{m_{\rm p}\over m_{\rm e}}{{(\theta_{\rm
e}+\theta_{\rm p})^{3/2}}\over{n_{\rm e}\sigma_{\rm
T}c\ln\Lambda}}\approx {{100}\over \tau_{\rm T}}t_{\rm cross},
\end{eqnarray}
where we
have assumed that the sum of $\theta_{\rm e}=kT_{\rm e}/m_{\rm e} c^2$
and $\theta_{\rm p}=kT_{\rm p}/m_{\rm p} c^2$ is of order unity. 
Provided that $T_{\rm e}\ll T_{\rm p},$ the coupling time to attain 
$T_{\rm e}$ 
\begin{equation}
t_{\rm e} =t_{\rm ep}{T_{\rm e}\over T_{\rm p}}\sim {0.1\over 
\tau_{\rm T}} t_{\rm cross}.
\end{equation} 

Thus if $T_{\rm e}\sim 10^9\K$ and $T_{\rm p}\sim 10^{12}\K$, $t_{\rm
e}\approx t_{\rm cross}/4,$ if $\tau_{\rm T}=0.4$, which means that $t_{\rm
e}\approx t_{\rm cool}$. This last condition is required for heating to
balance cooling for the electrons and determines $T_{\rm e}$.

A slab geometry enables $T_{\rm e}\sim 10^9\K$ for lower values of
$T_{\rm p}$. If, for instance, the thickness $h$ is one tenth of the
width $r$, the cooling time does not change, but the higher gas density
means that the coupling time decreases by the factor $h/r$. Moreover, if
the slab is favourably oriented so that it is perpendicular to the line
of sight, a luminosity variation can be observed on a time scale of $h/c$,
shorter than the one earlier assumed by the same factor. This enables
occasional variation to be detected on 100~sec (see e.g. Reynolds et al
1995). Thus $T_{\rm p}\approx 30({h/ r})({\ell/ \tau_{\rm T}})
T_{\rm e}.$

\subsection{Magnetic Field and Particle Energy Densities}

We take $\tau$ to be the Thomson optical depth
given by $\tau=n_e\sigma_{\rm T} h$. 
The size $h$ of an active region in the
corona can be constrained by observations of the shortest 
luminosity variability time scales $h/c \approx 100\s$ implying  
that $h\sim 3\times 10^{12}$cm.
For $\tau\sim 0.3$, $n\sim 1.5\times 10^{11}{\rm cm^{-3}}$.

Given that a fraction $\eta$ of the accretion
power
is dissipated in the corona $((1-\eta)$in the underlying disk), 
we have 
\begin{equation}
\eta L=4 \pi r^2 c \varepsilon_{\gamma}\sim 4\pi r^2 h
(\frac{d\varepsilon_{\rm m}}{dt})_-,
\end{equation}
where $\varepsilon_{\gamma}$ and $\varepsilon_{\rm m}$ are the photon
and magnetic energies respectively and $r$ is the radius of the
effective dissipation sheet of thickness $h$.  Now, $4 \pi r^2 h
({d\varepsilon_{\rm m}}/{dt})_- \sim 4 \pi r^2 v_{\rm dis}\times
B^2/8\pi$ so from eqn. 4 we have 
\begin{equation}
B^2/8\pi\sim
\varepsilon_{\gamma}(c/v_{\rm dis}).
\end{equation}
We assume that the dissipation velocity $v_{\rm dis}$ is a fraction
($f \approxlt 1$) of the Alfven speed $v_{\rm A}$, i.e. $v_{\rm dis}=f
B/\sqrt{4\pi \rho}$. In a reconnection site though, the fraction $f\sim
h/r$. From eqn. 6 we then have
\begin{eqnarray}
B=7.9\times 10^3\left(\frac{\eta L}{10^{43}{\rm erg\ s^{-1}}}\right)^{1/3}
\left(\frac{\rho}{10^{-13}{\rm g\ cm^{-3}}}\right)^{1/6} \nonumber 
\\ 
\times \left(\frac{h}{3\times 10^{12}{\rm cm}}\right)^{-1/3}
\left(\frac{r}{10r_g}\right)^{-1/3}\left(\frac{M}{10^7 M_\odot}\right)^{-1/3}
G, \end{eqnarray}
where $r_g\equiv GM/c^2$ and $M$ is the central (black hole) mass.
The characteristic Alfv\'en velocity would then
be $v_A \sim 0.2c$.

The above estimate uses only the energy transfer rates.
We now show that the assumption of equipartition between the magnetic
energy density $\epsilon_{\rm m}$, 
and proton energy density $\epsilon_{\rm p}=\epsilon_{\rm m}$
would likely underestimate the proton temperature.
In the steady state we know that 
$|(d\epsilon_{\rm m}/dt)_-|=|(d\epsilon_{\rm p}/dt)_+|=
|(d\epsilon_{\rm p}/dt)_-|$, and
the additional assumption of $\epsilon_{\rm m}=\epsilon_{\rm p}$
would mean that $t_{\rm ep}\equiv \epsilon_{\rm p}/|(d\epsilon_{\rm p}/dt)_-|
=\epsilon_{\rm m}/|(d\epsilon_{\rm m}/dt)_-|\equiv h/v_{\rm dis}$.
But using eq (3) with $h$ in place of $r$,  this would imply 
$v_{\rm dis}\sim \tau c /100\sim c/300$.  Since the dissipation 
physics of section 2 suggests $v_{\rm dis}\sim h/r$, 
and variability constraints imply $h/r\sim 1/10$, 
the assumption of equipartition between
the magnetic and proton energy densities would under estimate
the proton temperature by a factor $\sim 10$ compared to what 
we find in the next section.



We will see that the reason for this is that $t_{\rm ep}$ can be
one or two orders of magnitude longer than $t_{\rm e}$.
Therefore, before a steady state can ensue, the proton energy
density must build up to some critical value. The protons act as
a growing sea of magnetic energy until they can transfer
energy to electrons at the rate they gain energy from the magnetic field.  
Most of the energy in the active region thus resides in
the hot protons.  The least energy is contained in the electrons, 
whose energy density is less than the photon energy density $\epsilon_{\gamma}$ by the factor $t_{\rm cross}/t_{\rm
cool}$.
The long $t_{\rm ep}$
also means that 
although a flare may rise on a crossing time, its
decay time is uncertain.
The luminosity might drop gradually as the proton energy is exhausted or
abruptly if the region expands.


\section{A detailed model}
\subsection{Heating of electrons by ions}
We determine $T_{\rm e}$ by taking into account the
relevant cooling processes.  We solve
$(d\varepsilon_{\rm e}/dt)_-=(d\varepsilon_{\gamma}/dt)_+$, taking 
the transfer for Coulomb collisions between populations
of Maxwellian distributions 
using the Rutherford scattering cross-section (Stepney \& Guilbert 1983).
We show that typically, $T_{\rm e} \sim {\rm few}\ \times 10^9 \K$
and that Coulomb interactions are fast enough to explain the observed 
variability.  Future work should consider how
sensitive the result is to the presence non-thermal protons,
as the energization mechanisms of section 2 would likely provide them. 

We write
\begin{eqnarray}
(d\varepsilon_{\rm e}/dt)_- = \frac{3}{2}\,\frac{m_{\rm e}}{m_{\rm
p}}n_{\rm e}n_{\rm i}\sigma_{\rm T} c\frac{(kT_{\rm i}-kT_{\rm
e})}{K_2(1/\theta_{\rm e})K_2(1/\theta_{\rm i})}\ln{\Lambda} \nonumber
\\
\times \left[\frac{2{(\theta_{\rm e}+\theta_{\rm i})}^2 + 1}{\theta_{\rm e} + \theta_{\rm i}}K_1\left(\frac{\theta_{\rm e} + \theta_{\rm
i}}{\theta_{\rm e}\theta_{\rm i}}\right)+2 K_0\left(\frac{\theta_{\rm
e} + \theta_{\rm i}}{\theta_{\rm e}\theta_{\rm i}}\right)\right],
\end{eqnarray}
where ${\rm ln}\Lambda \sim 20$, is the Coulomb logarithm, the $K$'s
are the modified Bessel functions, and $\theta_{\rm e},\theta_{\rm
i}$ are the dimensionless electron and ion temperatures previously defined.
(Note that eqn. 6 is valid over all electron and proton temperatures). 

Electrons in the corona are cooled through inverse Compton scattering
off the soft photon field produced by the accretion disk, through
thermal synchrotron radiation and its Comptonization. Hence,
$(d\varepsilon_{\gamma}/dt)_+= q_{\rm
synch}+q_{\rm CS}+q_{\rm Comp}$.  The dominant term is usually
$q_{\rm Comp}=4 \theta_{\rm e}c \sigma_{\rm T} nU_{\rm
rad}$ where $\sigma_{\rm T}$ is the Thomson cross section and $U_{\rm
rad}=(1-\eta)L/4\pi R_{\rm disk}^2$.  The thermal synchrotron
emission rises steeply with decreasing frequency, becomes
self-absorbed below $\nu_{\rm c}$ and gives rise to a blackbody
spectrum..  The cooling is strongly peaked around $\nu_{\rm c}$ and can
be approximated by $q_{\rm synch}=(2\pi/3)m_{\rm
e}\theta_{\rm e} \nu_{c}^{3}/r$ (Narayan \& Yi 1995) where $\nu_{\rm
c}$ is obtained by equating the Rayleigh-Jeans emission to the
synchrotron one (Zdziarski 1985)
The Comptonization of the synchrotron radiation ($q_{\rm SC}$) is
calculated using Dermer and Liang (1991) approximate treatment.

Fig. 1 shows the electron temperature solution as a function of the
ion temperature when $(d\varepsilon_{\rm e}/dt)_-=(d\varepsilon_{\gamma}/dt)_+$
is solved. The figure illustrates how Coulomb heating of electrons in
the optically thin plasma of the localized regions naturally give rise
to a $T_{\rm e}$ consistent with X-ray observations of 
high energy spectral cut-offs in AGN.  Such
temperatures do not depend strongly on $T_{\rm p}$ as long
as protons are mildly relativistic. The $T_{\rm e}$ is
anticorrelated with $\tau$; as $\tau$ increases, the density $\rho$ and
consequently the magnetic field $B$ increase.  This implies
more synchrotron and Comptonised synchrotron cooling 
to balance an increased Coulomb coupling efficiency, resulting
in a lower $\theta_{\rm e}$.  The
temperature solution is not a strong function of the black
hole mass $M$ (see Fig. 1).

We note that for $\theta_{\rm e}$ and $\tau$ of
interest, electron-electron scattering through M\o ller cross section
competes with various loss mechanisms and in particular with Compton
losses in our case.  The question arises of whether
electrons can thermalize where the electron-electron relaxation
time scale $t_{\rm ee} = 4 \sqrt{\pi}\theta_{\rm e}^{3/2}/\ln\Lambda
(h/c\tau)\approxgt t_{\rm cool}$.  Ghisellini, Guilbert \& Svensson
(1988) though, found that cyclo/synchrotron self-absorption as
considered in our model, acts as a very efficient thermalizing
mechanism as long as the magnetic energy density dominates the
radiation energy density, which is a condition postulated here.
 We therefore expect electrons to be able to achieve a
Maxwellian distribution.

\begin{figure}
\caption{The dimensionless electron temperature $\theta_{\rm e}$ 
 plotted against ion temperature $T_{\rm i}$. The dashed line is for a
$\tau=0.1$ i.e. a lower density and lower magnetic field case.  The
synchrotron radiation and its comptonized emission become less
important cooling components in this case and the electron temperature
is slightly higher.  The two other lines are for $\tau=0.3$. The
dashed-dotted line is for a $M=10^{5}\Msun$ as opposed to the other
lines which correspond to typical AGN black hole mass of
$M=10^{7}\Msun$.  The electron temperature is
relatively independent of the proton temperature as long as the
latter become mildly relativistic.}
\end{figure}

\subsection{Time scale for electron heating}
In the previous section we have shown that  electron-ion Coulomb coupling 
can explain the values of $T_{\rm e}\sim {\rm few}\ \times 10^9$K 
inferred from observations. 
If particles are coupled only by two body collisions,
the protons must pass energy to electrons 
fast enough to explain the rapid
variability observed in X-ray sources.  
This requires $t_{\rm e \gamma}\equiv \varepsilon_{\rm \gamma}/(d\varepsilon_{\rm e}/dt)
= t_{\rm e}\varepsilon_{\rm \gamma}/\varepsilon_{\rm e}
  $  
 has to be less than the light crossing time $h/c$, where
$(d\varepsilon_{\rm e}/dt)_-$ is given by eqn. (8). Now since $n\sigma_{\rm
T}c=c\tau/h$, we have
\begin{equation}
t_{e}\approxlt \frac{2}{3}\frac{m_{\rm p}}{m_{\rm e}}\frac{1}{\ln
\Lambda}\frac{T_{\rm e}}{T_{\rm i}-T_{\rm e}} f(\theta_{\rm e},
\theta_{\rm i})\frac{h}{c\tau}=\Psi \frac{h}{c\tau}
\end{equation}
where $f(\theta_{\rm e},\theta_{\rm i})$ is the function of
$\theta_{\rm e},\theta_{\rm i}$ in eqn. (8) and $t_{\rm e}$ is expressed
as a fraction ($\Psi$) of the time scale on which photons escape
($h/c$) a dissipation region. In eqn. (12) $\Psi \approxlt 1/10$ and
almost invariant of $T_{\rm i}$ (for $T_{\rm i} \approxgt 10^{12}\K$
(see Fig. 2), so that for a typical $\tau \approxlt 0.4$,
$t_{e}\approxlt h/c$, 
consistent with $t_{\rm e\gamma }/t_{\rm e}\sim  c/h\sim t_{\rm e}
\varepsilon_{\rm \gamma}/\varepsilon_{\rm e}$ equal
to a few.  Variability constraints are
thus consistent with ion-electron Coulomb
collisions being the dominant process for electron heating.
\begin{figure}
\caption{The function $\Psi$ plotted against ion temperature $T_{\rm i}$
As indicated in the figure, the solid line is for $\theta_{\rm e}=0.4$
and the the dashed one for $\theta_{\rm e}=0.6$. For $T_{\rm
i}\approxgt 10^{12}$ the function $\Psi$ becomes virtually constant so
as to give a ratio $\Psi/\tau \approxlt 1$ for typical electron
temperature and therefore an electron heating time
less than or 
comparable to the light crossing time for the active current sheet in
the corona.}
\end{figure}

\section{Discussion}
We have demonstrated that an electron temperature of a few $\times 10^9
K$, determined from interpretation of X-ray observations, can be
very naturally explained by an optically thin two-temperature plasma corona
model with $T_{\rm i} >> T_{\rm e}$ if Coulomb scattering is the
dominant source of electron heating.  
If magnetic dissipation in the
corona preferentially energizes the ions, then 
$T_{\rm i}$ will naturally exceed $T_{\rm e}$.  
The efficient cooling of electrons then leads a $T_{\rm e}$ 
being locked at the required values.  
The result of $T_e\sim 10^9$K is robust 
because $T_e$ is relatively insensitive to $T_p$ when
the latter exceeds $10^{12}$K.

We have also shown that the time scale for electron
heating by Coulomb coupling is consistent with the
constraints required by observations of fast variability in AGN.
Though we have considered primairly thermalized ions, 
the actual dissipation may lead to non-thermal ion populations.  
This would not likely effect 
$t_{\rm e}$ strongly, but the effect should be considered in future work.
Regardless of the proton distribution, the electron distribution would 
still become Maxwellian.

Pronounced variability suggests that the corona likely dissipates
in large flares, rather than in a large number of small dissipation
sites.  
Since the typically observed  variabilities range between $100-1000$sec
and not much shorter, the effective aspect ratio of the dissipation region
can be constrained to satisfy $h/r \sim 100/1000=1/10$.  
We can speculate that a small
number of operating current sheets would mean that this aspect ratio
characterizes the dimensions of a typical reconnection region itself. 
Such a thick aspect ratio might suggest that Petschek rather than 
Sweet-Parker type reconnection is occurring.  
In any case, the physics of dissipation could in principle constrain
this value. Future work is needed to fully address the dissipation
physics.  

The model predicts that the proton energy 
probably dominates the magnetic energy density, but both are much larger than
the photon and electron energy densities.  
The electron energy density is less than the 
photon energy density by a factor of the cooling time over the
light crossing time.  These differences in energy contents
highlight the fact that balancing the rates of energy transfer
does not necessarily imply equipartition between any energy densities.

Throughout this paper, we have assumed that the only coupling between
ions and electrons is via Coulomb collisions which leads to 
a two-temperature plasma. However, Begelman
\& Chiueh (1989, hereafter BC) have discussed a plasma instability
which may lead to thermal coupling of ions and electrons in a
two-temperature flow ($T_{\rm i} > T_{\rm e}$) on a time scale shorter
than the Coulomb coupling time.  The instability
may grow in regions with a high level of small
scale MHD turbulence that have a plasma $\beta$ parameter 
$\beta\equiv 8\pi nk T_{\rm i}/B^2 \approxgt 1$.  We can estimate 
the rate of electron heating via the BC mechanism using the specific parameters
which characterise our solution and compare it to the rate of energy
transfer via Coulomb collisions (eqn. 8).  
For simplicity we choose
$T_{\rm i}=10^{12}$. According to BC, the rate of heating of electrons
when their instability is fully developed is given by
\begin{equation}
\left(\frac{d\varepsilon}{dt}\right)_{\rm BC} \approx f \frac{m_{\rm i}v_{\rm ti}^3}{\lambda_{\rm Di}}\left(\frac{v_{\rm A}}{c}\right)^2\left(\frac{\lambda_{\rm ci}}{L}\right)^{9/2},
\end{equation}
where $f$ is the filling factor of the plasma where the strong
turbulence needed for the instability is present, $m_{\rm i}$ is the
ion mass, $v_{\rm ti}$ is the ion thermal velocity, $L$ is the local
density gradient scale and $\lambda_{\rm Di}$ is the ion Debye length
and $\lambda_{\rm ci}$ its respective gyroradius (where $\lambda_{\rm
ci}=(\beta/\theta_{\rm i})\lambda_{\rm Di}$).  We find that for
$v_{\rm A}/c\sim 1/5$ and for $L$ satisfying the condition for the BC
instability to operate (eqn. 5.6 in BC paper) with our choice of $\beta$,
we have that
\begin{equation}
\frac{(d\varepsilon/dt)_{\rm BC}}{(d\varepsilon/dt)_{\rm e}}\approxlt 0.3 f
\end{equation}
This shows that even in the occurrence of the BC instability, the
Coulomb heating usually dominates for the choice of parameters of
interest in our solution.

Finally, since the coronal plasma discussed herein has 
very high ion temperatures, e.g. $T_{\rm i} \approxgt 10^{12}$ we
expect collisions between protons in the high energy tail of the
distribution to produce pions \footnote{This process cools the protons
without heating the electrons and for ion-temperature of interest it
would not cool the ions faster than Coulomb collisions.}. These would
then decay into high energy $\sim 70 \MeV \gamma$--rays. Such
$\gamma$--ray emission would constitute a testable feature for this
model. Power-law or Maxwellian proton (or combination of the two)
distributions could be distinguished by the different $\gamma$-ray
spectra they predict (Dermer 1986).

\section*{Acknowledgements}
For financial support, we acknowledge PPARC and Trinity College,
Cambridge (TDM), the Royal Society (ACF).  We thank M. Rees for discussion.

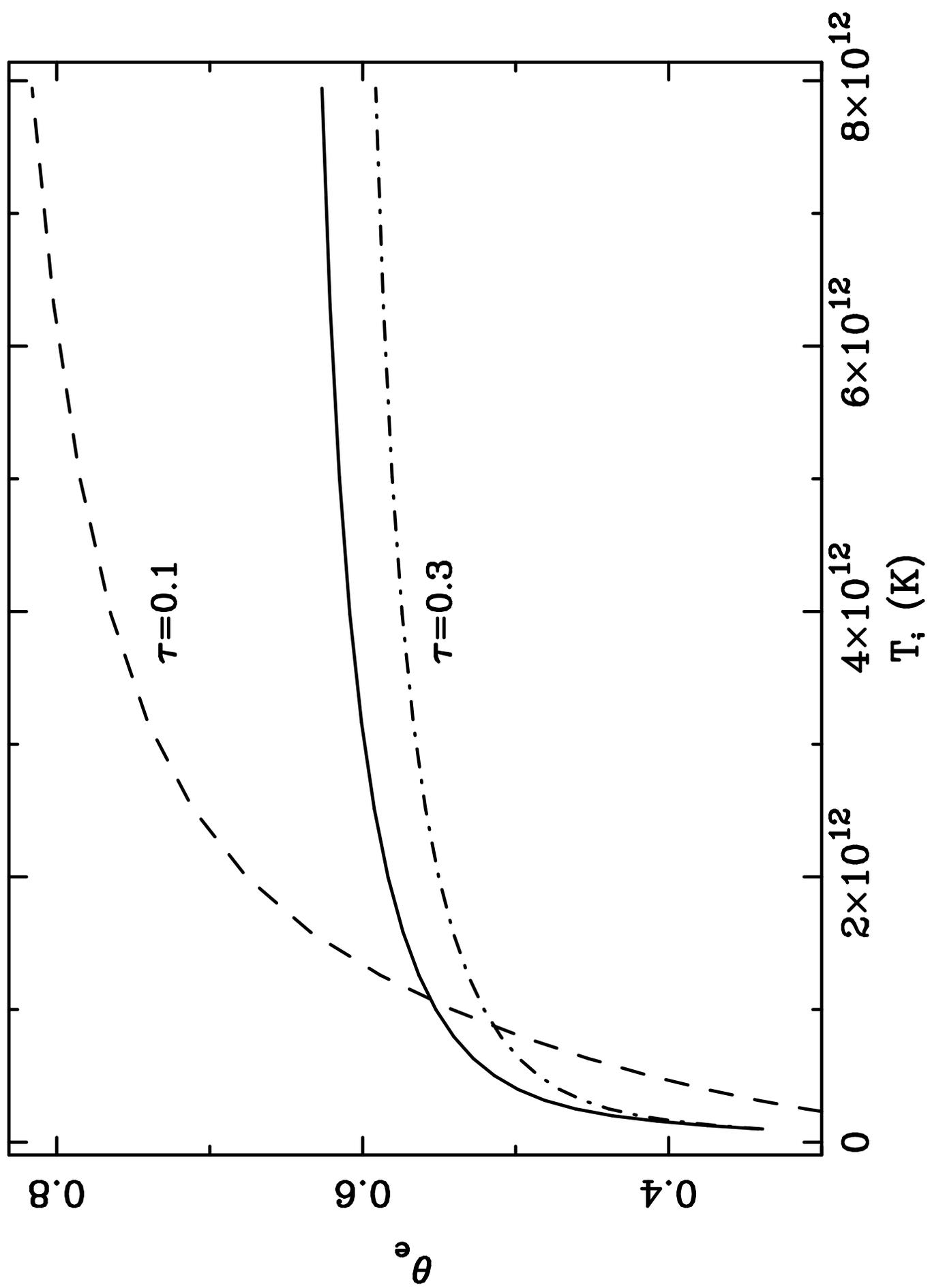
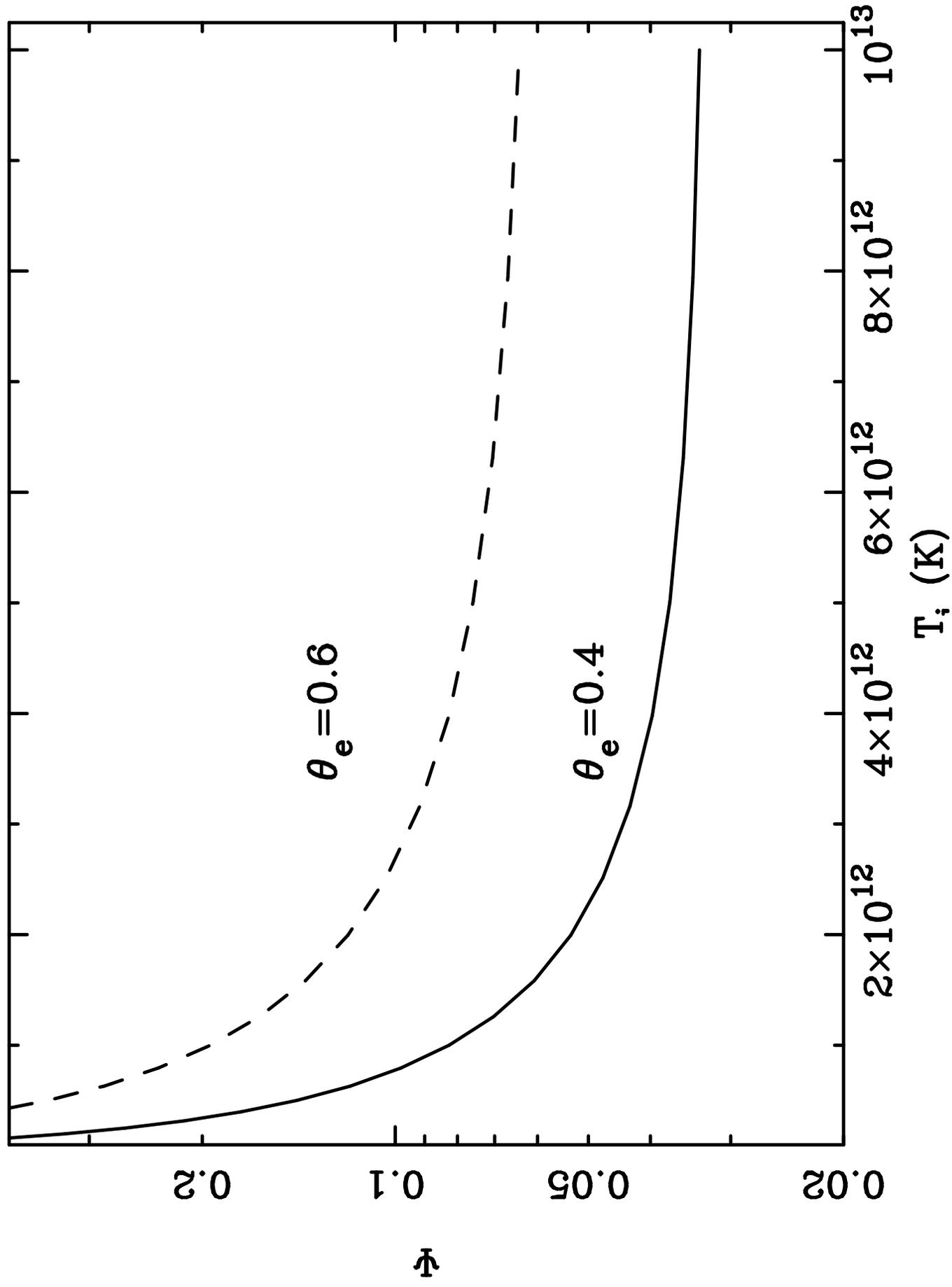
\end{document}